\documentclass[apj]{emulateapj}
\usepackage{epsfig}
\usepackage{amsmath, amssymb,bm}
\usepackage[varg]{txfonts}
\usepackage{color}

\begin{document}
\label{firstpage}

\title[The Solar System as an Exoplanetary System]{The Solar System as an Exoplanetary System}

\author{Rebecca G. Martin\altaffilmark{1}} 
\author{Mario Livio\altaffilmark{2}}
\affil{\altaffilmark{1}Department of Physics and Astronomy,
  University of Nevada, Las Vegas, 4505 South Maryland Parkway, Las
  Vegas, NV 89154, USA }
\affil{\altaffilmark{2}Space Telescope Science Institute, 3700 San
  Martin Drive, Baltimore, MD 21218, USA }

\begin{abstract} 

With the availability of considerably more data, we revisit the question
of how special
our Solar System is, compared to observed exoplanetary systems. To this
goal, we employ a mathematical transformation
that allows for a meaningful, statistical comparison. We find that the
masses and densities of the giant planets in our Solar System
are very typical, as is the age of the Solar System. While the orbital
location of Jupiter is somewhat of an outlier, this
is most likely due to strong selection effects towards short-period
planets. The eccentricities of the planets in our Solar System
are relatively small compared to those in observed exosolar systems, but
still consistent with the expectations for an 8-planet system (and could,
in addition, reflect a selection bias towards high-eccentricity planets).
The two characteristics of the Solar System that we find to be most
special are the lack of super-Earths with orbital periods of days to
months
and the general lack of planets inside of the orbital radius of Mercury.
Overall, we conclude that in terms of its broad characteristics
our Solar System is not expected to be extremely rare, allowing for a
level of optimism in the search for extrasolar life.

%With the availability of considerably more data, we revisit the
%question of how special our Solar System is compared to observed
%exoplanetary systems. To this goal, we employ a mathematical
%transformation that allows for a meaningful, statistical
%comparison. {\bf However, strong observational biases that still
%  affect the sample of observed exoplanets probably make some
%  properties of the Solar System appear more special than they truly
%  are. } We find that the masses and densities of the giant planets in
%our Solar System are very typical. {\bf However, the lack of a
%  super-Earth in the Solar System could be unusual since more than
%  half of all the observed exoplanetary systems appear to contain
%  one. } The orbital location of Jupiter remains somewhat of an
%outlier, but this is most likely due to {\bf very strong selection
%  effects towards short--period planets}. The relative depletion of
%mass in the terrestrial region is probably the Solar System's most
%distinctive characteristic.  The eccentricities of the planets in our
%Solar System are relatively small compared to those in observed
%exosolar systems.  {\bf However, this could be due to section biases
%  towards high eccentricity planets.} The low eccentricities are
%consistent with the expectations for an 8-planet system.  Overall, we
%conclude that in terms of its broad characteristics our Solar System
%is not expected to be extremely rare, allowing for a level of optimism
%in the search for extrasolar life.
\end{abstract} 
 
 
\section{Introduction} 
 
The discovery of thousands of extrasolar planets and planet candidates
in recent years \citep[see, e.g.,][and references therein and see
  exoplanet.org for a complete
  list]{Wrightetal2011,Batalha2014,Rowe2014,Lissauer2014, Han2014}, coupled with
the rapidly increasing interest in the potential existence of
extrasolar life, raise again in a big way the question of whether or
not our Solar System is special in any sense. More specifically, we
are interested in understanding whether the planetary and orbital
properties in our Solar System are typical or extremely unusual
compared to those of extrasolar planets.
 
The Solar System contains eight planets and two main belts (the
asteroid belt and the Kuiper belt).  While tens of debris disks and
warm dust belts (similar perhaps to the Solar System's asteroid belt)
have been observed and resolved, belts with dust masses as low as
those in the Solar System would currently be undetectable in
extrasolar systems \citep[e.g.][]{Wyatt2008, Panic2013}. Consequently,
we can quantitatively assess in detail how special the Solar System
is, only on the basis of its planetary components and properties such
as its age and metallicity. However, there are now hundreds of
unresolved debris disk candidates \citep[e.g.][]{Chen2014}. Of these, about
two thirds of the systems are better modelled by a two component dust
disk rather than a single dust disk. The two temperature components
likely arise from two separate belts \citep{Grant2014}. Thus, the two
belt configuration of our own Solar System is plausibly fairly
typical.

\cite{Beer2004} made an initial attempt to explore to what extent
Jupiter's periastron could be considered atypical compared to those of
the giant planets known at the time.  Their analysis, however,
included only fewer than 100 exoplanets, most of which had been
detected via radial velocity measurements.  Consequently, selection
effects dominated their conclusions---a possibility fully acknowledged
by the authors. In the present work we re-examine the question of how
special the Solar System is. In Section~\ref{special} we use the much
larger currently available database to consider the planetary orbital
parameters.  We identify the semi--major axis of the innermost
  planet as the most discrepant characteristic of the solar system and
  the low mean eccentricity as being somewhat special. In
Section~\ref{prop} we compare the masses and densities of the planets
in our Solar System with those in exosolar systems.  While the
  lack of a super--Earth in the Solar System is somewhat unusual, we
  argue that none of the characteristics identified make the Solar
  System very special.  We discuss potential implications of our
results in Section~4.  Some of the apparent differences between the
  Solar System and exoplanetary systems continue to be driven by
  strong selection effects that affect the sample.  We draw our
conclusions in Section~5.
 
\section{Planetary Orbital Properties} 
\label{special}

We first consider the statistical distribution of orbital separation
and eccentricity of the observed planetary orbits. To allow for a more
meaningful quantitative analysis, we perform a \cite{BoxCox}
transformation on the data. This transformation makes the data closer
to a normal distribution so that we can more accurately evaluate
properties such as the mean and the standard deviation.  The
transformation takes a skewed data set to approximate normality. It is
based on the geometric mean of the measurements and is independent of
measurement units.  It is possible to do a multivariate Box-Cox
transformation \citep[e.g.][]{Velilla1993}. However, because of the
selection effects associated with different parameters we choose to
consider each parameter separately.

We transform the data with the function
\begin{align} 
y_{\lambda}(a)= 
\begin{cases} 
        \begin{array}{lcl} 
        \frac{a^\lambda-1}{\lambda} & {\rm if}~~ \lambda \ne 0 \\  
         \log a &{\rm if}~~\lambda =0,
        \end{array} 
  \end{cases} 
\end{align} 
where $a$ is the parameter we are examining, such as the semi-major
axis or eccentricity, and $\lambda$ is a constant that depends upon
the original distribution, that we discuss below. The maximum
likelihood estimator of the mean of the transformed data is
\begin{equation} 
\overline{y}_\lambda=\sum_{i=1}^n\frac{y_{\lambda,i}}{n}, 
\end{equation} 
where $y_{\lambda,i}=y_\lambda(a_i)$ and $a_i$ is the $i$-th
measurement of a total of $n$. Similarly, the maximum likelihood
estimator of the variance of the transformed data is
\begin{equation} 
s_\lambda^2=\sum_{i=1}^n \frac{(y_{\lambda,i}-\overline{y}_\lambda)^2}{n}. 
\end{equation} 
We choose $\lambda$ such that we maximise the log likelihood function 
\begin{equation} 
l(\lambda)=-\frac{n}{2}\log(2\pi)-\frac{n}{2}-\frac{n}{2}\log s_\lambda^2+(\lambda-1)\sum_{i=1}^n\log a_i. 
\end{equation} 
This new distribution, $y_\lambda(a)$, will be an exact normal
distribution if $\lambda=0$ or $1/\lambda$ is an even integer.
 
We can measure how well the transformed distribution compares to a
normal distribution with two parameters.  The skewness is
\begin{equation} 
S=\sum_{i=1}^n\frac{1}{n}\left(\frac{y_{\lambda,i}-\overline{y}_\lambda}{\sigma}\right)^3, 
\end{equation} 
where $\sigma$ is the standard deviation of the distribution and thus
we take $\sigma=s_\lambda$. The skewness measures the asymmetry of the
distribution, a positive number implying the right hand tail of the
distribution is longer, and a negative number that the left hand tail is
longer. Furthermore, we can compare the kurtosis,
\begin{equation} 
K=\sum_{i=1}^n\frac{1}{n}\left(\frac{y_{\lambda,i}-\overline{y}_\lambda}{\sigma}\right)^4 -3. 
\end{equation} 
This is a measure of the ``peakedness'' of the distribution and the
heaviness of the tails. A normal distribution has a kurtosis value of
zero. A positive value means that the distribution is tightly peaked
but the tails are broad, and vice versa for a negative value. In the
following subsections we take the samples of the eccentricity and
semi-major axis of the observed exoplanets and compare them to the
planets in our own Solar System.

\begin{figure*} 
\begin{centering} 
\includegraphics[width=8.5cm]{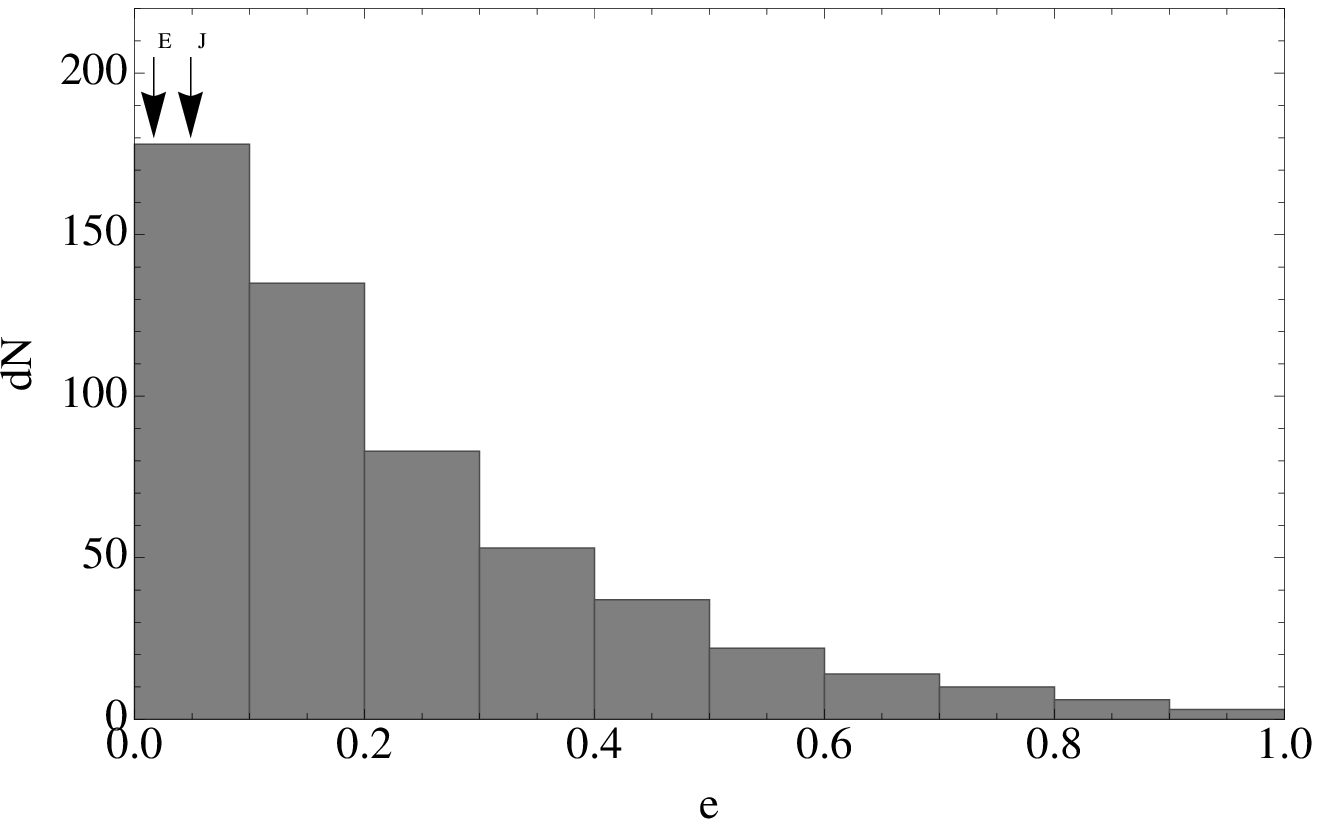} 
\includegraphics[width=8.5cm]{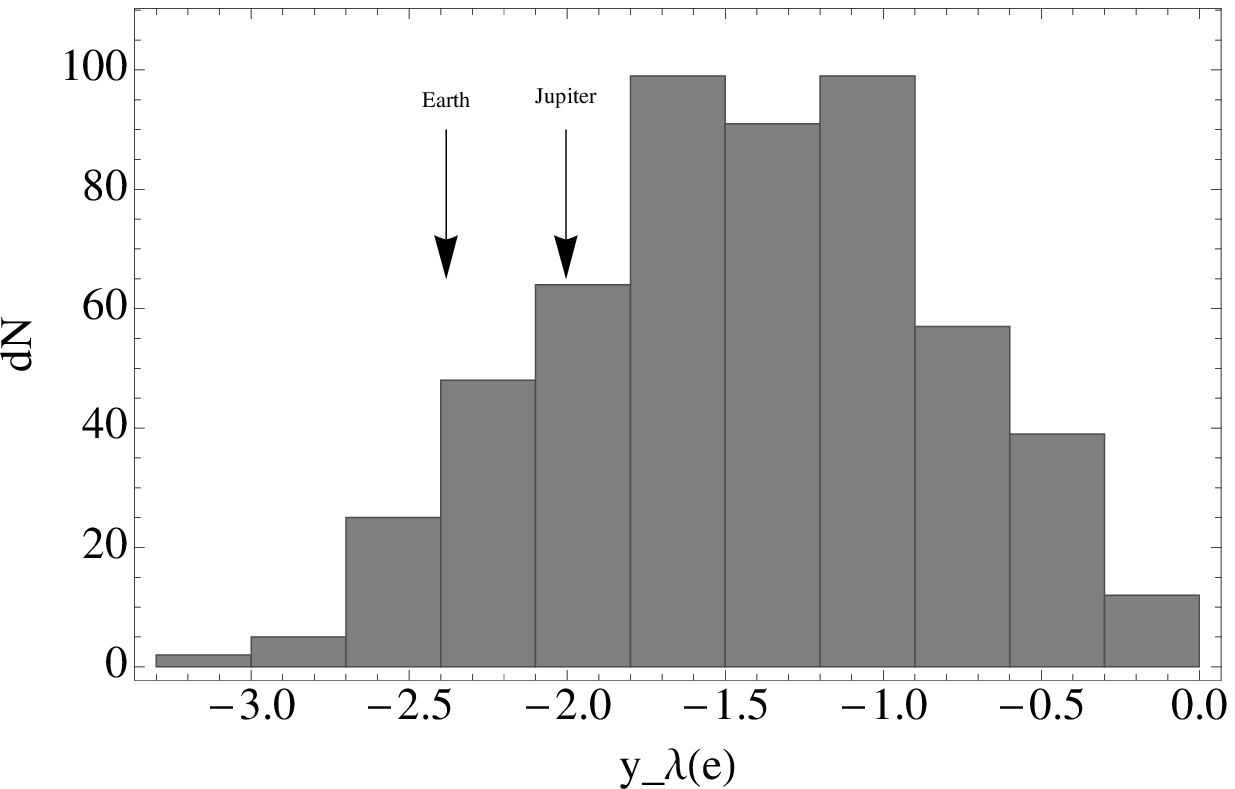} 
\end{centering} 
\caption{Left: Eccentricity distribution for all observed exoplanets 
  with a measured orbital eccentricity.  Right: Box-Cox transformed 
  distribution of exoplanet eccentricities. The total number of 
  exoplanets is 539. } 
\label{ecc} 
\end{figure*}

\subsection{Eccentricity} 
 
There are a total of 539 exoplanets with a measured eccentricity.  In
the left hand panel of Fig.~\ref{ecc} we show the eccentricity
distribution for this sample. We find the maximum of the log
likelihood function to be when $\lambda=0.30$. The right hand panel of
Fig.~\ref{ecc} shows the histogram of the Box-Cox transformed data.
For our data, we find the skewness to be $S=-0.065$.  For a normal
distribution we expect the magnitude of the skewness to be up to
$\sqrt{6/n}=0.11$. Thus the data are not heavily skewed. The kurtosis
is $K=-0.53$ whereas for a normal distribution we would expect the
magnitude to have values up to $\sqrt{24/n}=0.21$. Thus, the data have
a slightly large kurtosis, or the distribution is not so tightly
peaked as a normal one. For the transformed data, the mean is $-1.42$
and the standard deviation is $0.60$. Jupiter lies at $-0.97\sigma$
from the mean. Similarly the Earth lies at $-1.60\sigma$. Thus, the
eccentricities of the planets in our Solar System (that range from
Venus at $e=0.0068$ to Mercury at $e=0.21$) are all relatively small
compared to those of exoplanets, but not altogether significantly
different.

Recently \cite{Limbach2014} considered how the mean eccentricity of
planets in a system is correlated with the number of planets in the
system. They found a strong anti-correlation of eccentricity with
multiplicity in systems observed by radial velocity. An extrapolation
of their relation up to 8 planets fits well with the mean eccentricity
observed in the Solar System.  Furthermore, \cite{VanEylen2015} used
the Kepler exoplanets with asteroseismically determined stellar mean
densities to derive a rather low eccentricity distribution of the
multi--planet Kepler systems. Thus, while the eccentricities in our
Solar System are low, those may be expected in a system with so many
planets.

There is some bias in the exoplanet eccentricity data. For the RV
planets, the best fit eccentricity is biased upwards from the true
value leading to a reduced number of systems with a low eccentricity
\citep[e.g.][]{Shen2008,Hogg2010,Zakamska2011, Moorhead2011}. However,
the detection efficiency decreases only mildly with increasing
eccentricity because despite being more difficult to detect, they have
a larger RV amplitude for a fixed planet mass and semi-major axis
\citep{Shen2008}. While planets found with the transit method require
follow up observations (for example, with RV) to determine the
eccentricity, the distribution of eccentricities is consistent with
those of the RV planets \citep{Kane2012}. Without the bias, the
eccentricities of the planets in our Solar System would appear to be
less special.

\subsection{Semi-Major Axis} 
\label{a}

\begin{figure*} 
\begin{centering} 
\includegraphics[width=8.5cm]{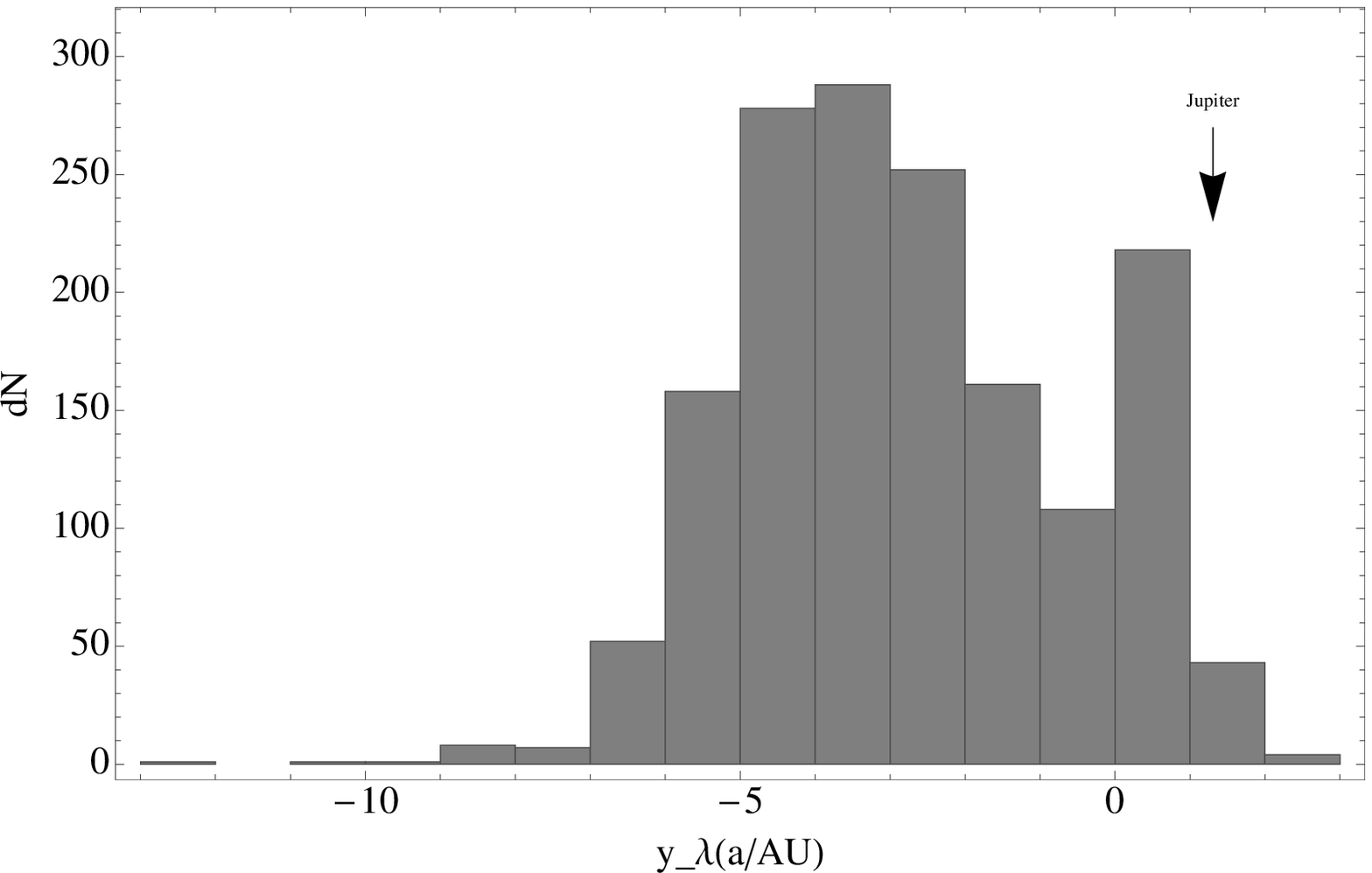}  
\includegraphics[width=8.5cm]{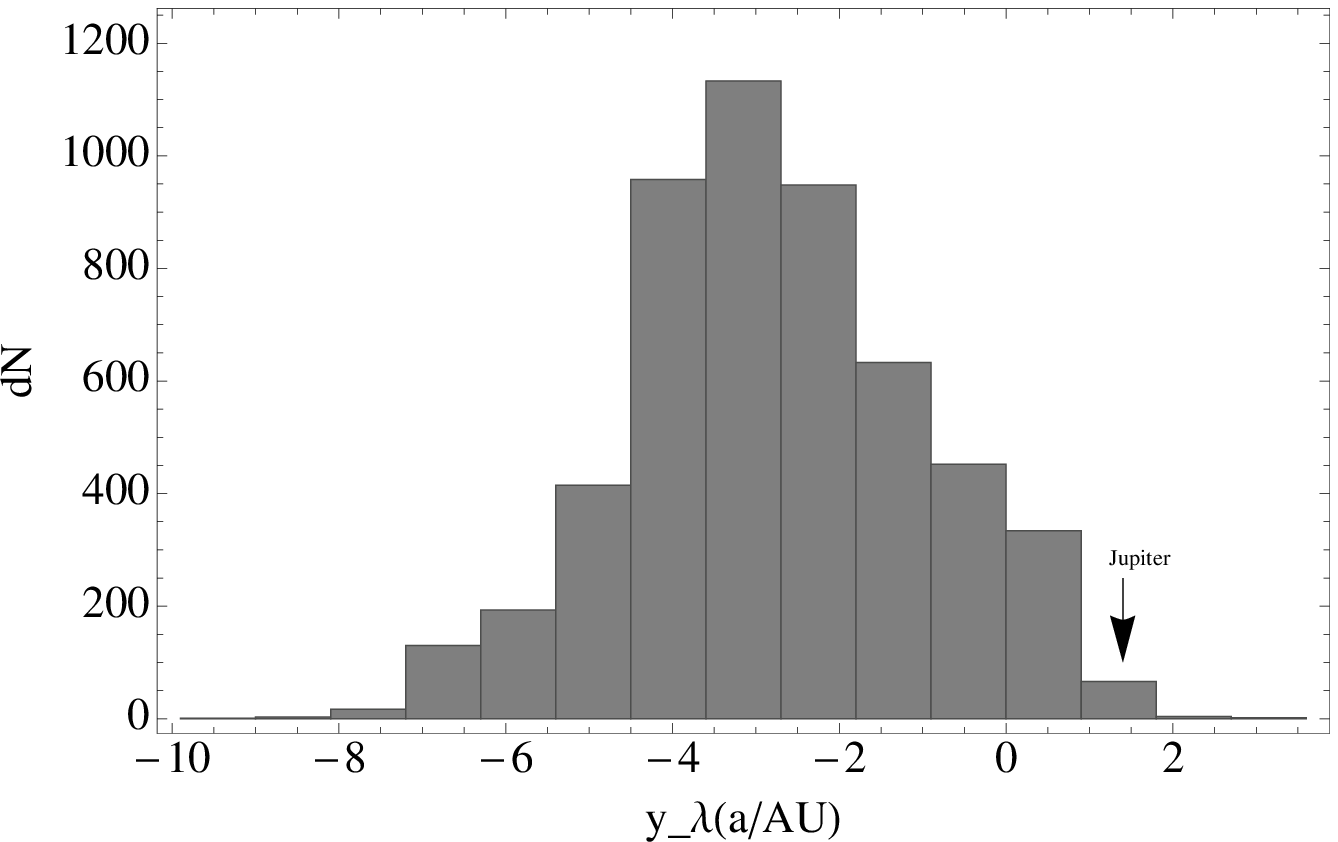} 
\end{centering} 
\caption{Box-Cox transformed distribution of exoplanet semi-major
  axis. Left: Only including confirmed exoplanets. The total number of
  exoplanets is 1580. Right: Including all planet candidates. The total
  number of exoplanets is 5289. }
\label{data} 
\end{figure*}

To date, there are a total of 1580 planets with a determined
semi-major axis.  This increases up to a total of 5289 if we include
planet candidates from Kepler. The false positive rate of the Kepler
exoplanets is low, especially for multi planet systems and the
non--giant planets \citep[e.g.][]{Desert2015} and thus we consider
this much larger dataset also.  We find the maximum of the log
likelihood function to be when $\lambda=-0.29$ ($\lambda=-0.20$,
including planet candidates). Fig.~\ref{data} shows the histogram of
the Box-Cox transformed data.  The left hand panel includes only
confirmed exoplanets.  For this data, we find the skewness to be
$S=0.11$.  This is only slightly higher than the upper value expected
for a normal distribution of $\sqrt{6/n}=0.062$.  We find the kurtosis
to be $K=-0.48$ whereas for a normal distribution we would expect
magnitudes smaller than $\sqrt{24/n}=0.12$. Thus, the data have a
large kurtosis. For the transformed data, the mean is $-2.77$ and the
standard deviation is $2.16$. Thus, Jupiter lies at $1.9\sigma$.  The
right hand panel of Fig.~\ref{data} repeats this analysis but includes
unconfirmed Kepler planets. The skewness for this is small at
$S=0.022$ (expected magnitude less than $0.034$) and the kurtosis is
much smaller also, $K=-0.18$ (expected magnitude less than
$0.067$). Jupiter lies at $2.4\sigma$, suggesting on the face of it
that Jupiter is rather special. However, as we explain below, this is
most likely a result of selection effects.

The majority of the planets in this distribution have been found by
transit methods. The planet with the largest semi-major axis found by
this method is only at $0.996\,\rm AU$ \citep[the planet is KIC
  11442793~h,][]{Cabrera2014}.  Kepler, is thought to be complete only
for planets at least as large as the Earth and for orbital periods up
to a year \cite[e.g][]{Winn2015}. Microlensing surveys preferentially
find planets at radial distances of a few AU from their host star,
that is often an M dwarf \citep[e.g][]{Gould2010,Cassan2012}. This
scale is dictated by the size of the Einstein ring radius around the
lensing star. The radial velocity method has found planets in the
range $0.01$ to $5.83\,\rm AU$ \citep[e.g.][]{Bonfils2013}. Direct
imaging can detect planets at much larger distances
\citep[e.g.][]{Marois2008,Lafreniere2010}, but so far only 8 planets
have been detected by the method.

\begin{figure} 
\begin{centering} 
\includegraphics[width=8.5cm]{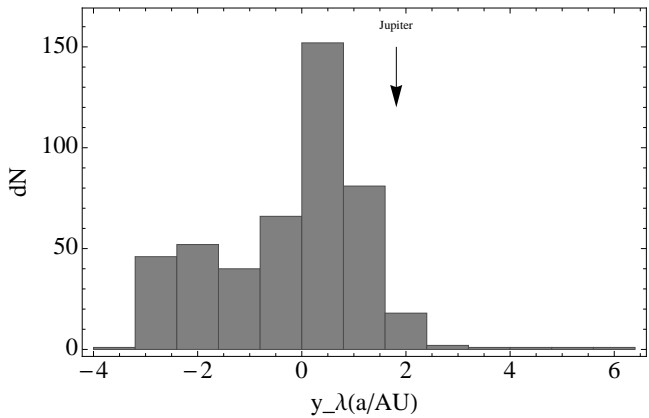} 
\end{centering} 
\caption{Box-Cox transformed distribution of exoplanet semi-major axis
  not including planets found by the transit method. The total number
  of exoplanets is 473. }
\label{datano} 
\end{figure}

In order to test the possibility that Jupiter's outlier status is
largely due to selection effects, we repeated the analysis but removed
planets found by the transit method. There remain $473$ planets in the
sample.  We find the maximum of the log likelihood function to be when
$\lambda=0.11$.  Fig.~\ref{datano} shows the histogram of the
transformed data.  For our data, we find the skewness to be $S=0.015$.
This is less than the value expected for a normal distribution of
$\sqrt{6/n}=0.11$.  For our data we find a high kurtosis of $K=0.71$
whereas for a normal distribution we would expect values with
magnitude less than $\sqrt{24/n}=0.23$.  For the transformed data, the
mean is $-0.22$ and the standard deviation is $1.41$. Thus, Jupiter
lies at $1.44\sigma$ and continues to be somewhat of an outlier, but
the trend suggests that this is most likely still due to selection
effects. The fact that direct imaging repeatedly reveals planets at
separations much larger than Jupiter's also may indicate that the
current relative dearth of planets at large separations could be due
to selection biases but more complete observations are required to
test this possibility.

We should also note that \cite{Beer2004} used only the planet with the
largest velocity semi-amplitude in each observed system in their
plots. However, they also performed the analysis with the most massive
planet in each system and again with all the planets. They reported no
difference in the significance of Jupiter as an outlier. In 2004, they
found that Jupiter was at $2.3\sigma$ and half a sigma from its
nearest neighbour. We find that Jupiter is not such an outlier as it
was with the much smaller data set in 2004, and selection effects
continue to affect the distribution. This analysis should again be
repeated once we have more reliable observations around the orbital
radius of Jupiter.

\subsection{Inner Solar System}

Currently, our inner Solar System appears to be rather special
compared to observations of exoplanet systems. The inner edge of our
Solar System is at the orbit of Mercury at $0.39\,\rm AU$, while
exoplanetary systems are observed to habor planets much closer to
their star. We find that Mercury lies at $0.78\,\sigma$ above the
mean, while the Earth is at $1.28\,\sigma$ above the mean of the
distribution of confirmed exoplanet semi-major axes (as shown in the
left hand panel of Fig.~\ref{data}).  When we include all of the
Kepler candidates (right hand panel of Fig.~\ref{data}), these
increase to $0.98\sigma$ for Mercury and $1.57\sigma$ for the
Earth. Thus, all of the planets in our Solar System have orbital
semi-major axes that are larger than the mean observed in exoplanetary
systems. However, it is possible that this could be the result of
selection effects as it is easier to find planets in this region, if
they are there.  In terms of the radial location of observed
exoplanets, the lack of close--in planets in our Solar System is the
parameter that makes our Solar System most special. We should note
though that if we use only the planets found by methods other than the
transit, then Mercury is at $-0.48\sigma$ and the Earth is at
$0.16\sigma$. Consequently it is difficult to say how significant this
discrepancy is.

\cite{Batygin2015} suggested that the migration of Jupiter and Saturn
into the terrestrial planet forming region of our Solar System (down
to $a\approx 1.5\,\rm AU$) led to the depletion of mass in
$a<0.39\,\rm AU$. The planets are then thought to migrate outwards to
their current location \citep[see also][]{Walsh2011}. However,
whatever the formation mechanism for the giant planets in our Solar
System might have been, it is not thought to have been specific to our
Solar System, and thus neither is a depleted inner Solar System. We
discuss this point further in Section~\ref{discussion}.

\subsection{Migration}

Finally in this Section, we note that migration of planets through the
protoplanetary disk or planet-planet interactions or secular
interactions of a binary star could affect these distributions,
especially that of the semi-major axes. For example, it may be
theoretically impossible for Jupiter mass planets to form at the
radial location of hot--Jupiters
\citep[e.g.][]{Bodenheimer2000}. Instead, they are supposed to form
outside of the snow line and migrate inwards through the
protoplanetary disk before the latter disperses
\citep[e.g.][]{Goldreich1980}.  Migration could also occur by the
Kozai--Lidov mechanism increasing the eccentricity of the planet
followed by tidal circularization \citep{Kozai1962, Lidov1962, Wu2003,
  Takeda2005,Nagasawa2008,Perets2009}. Evidence obtained by the
Rossiter--McLaughlin effect suggests that some fraction of the hot
Jupiters may have been produced through dynamical interactions
\citep[see
  e.g.][]{Winn2005,LubowIda2010,Triaudetal2010,Albrechtetal2012}.
Hot--Jupiter planets dominated the initial planet discoveries because
they are large and close to their host star. However, we now know that
they are quite rare and orbit only about one percent of solar type
stars \citep[e.g.][]{Wright2012}.

Opinions vary on whether in the Solar System Jupiter has significantly
migrated. On one hand \cite{Morbidelli2010} suggested that Jupiter did
not migrate much from its formation location, and on the other,
\cite{Walsh2011} proposed that the low mass of Mars could be explained
by gas-driven early migration of Jupiter. \cite{Batygin2015} further
suggested that this formation process could explain the lack of
objects in our inner Solar System. \cite{Armitage2002} assumed that
planets form constantly at a radius of $5\,\rm AU$ and found
theoretically that about $10$ to $15\%$ of systems will have a Jupiter
mass planet that does not migrate significantly. However, this
conclusion will be affected by the presence of a dead zone \citep[a
  region of the disk with no turbulence,
  e.g.][]{Gammie1996,Martinetal2012a,Martinetal2012b} that may slow or
halt migration altogether. Furthermore, the inner and outer edges of a
dead zone may act as planet traps that stop migration
\citep[e.g.][]{Hasegawa2011, Hasegawa2013}. Thus, the distribution of
planet semi-major axes definitely does not represent the initial
distribution at the time of planet formation.

Given that Jupiter mass planets are thought to form in the vicinity of
Jupiter's current radial location, theory suggests that the radial
location of Jupiter is not particularly special. The current
observational bias towards planets that are close to their host star
means that it is easier to find planets that have migrated inwards,
rather than those that have not, or even those that may have migrated
outwards. We expect in the future that with more complete observations
of Jupiter mass planets at Jupiter's radial location we will be able
to constrain the migration mechanisms and uncover how special Jupiter
really is for its small (net) distance of migration.

\section{Planetary Properties} 
\label{prop}

In this Section we consider how the masses and densities of the
planets in our Solar System compare to those in exosolar systems. In
this context, we discuss also the potential significance of the lack
of a super-Earth in our Solar System.

\subsection{Planet Masses}
 
\begin{figure} 
\begin{centering} 
\includegraphics[width=8.4cm]{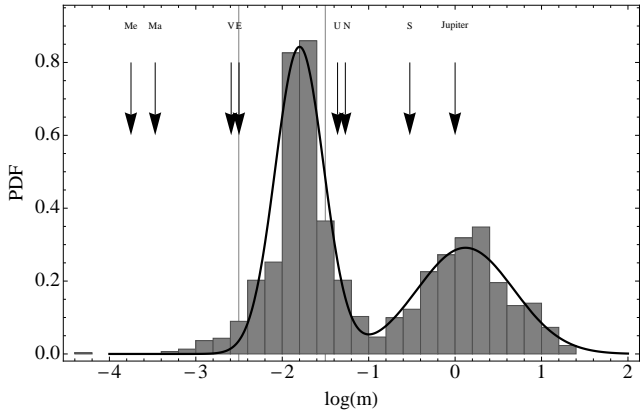} 
\end{centering} 
\caption{Exoplanet mass distribution. The arrows show the masses of the 
  planets in our Solar System. The vertical lines shows range of 
  planets that are considered to be super--Earths, the lower limit is 
  the mass of the Earth and the upper limit is the lower limit to the 
  mass of a giant planet at $10\,M_\oplus$. The total number of 
  exoplanets is 1516. } 
\label{mass} 
\end{figure}

Fig.~\ref{mass} shows the distribution of the approximate masses of
the exoplanets that have been observed to date\footnote{The masses for
  planets that have been observed by the Doppler method are $M_{\rm
    p}\sin i$, where $i$ is the orbital inclination. For directly
  imaged planets, the mass is predicted by theoretical models of the
  planets' evolution. For the planets that have been observed by
  microlensing, it's the ratio of the planet to star mass that is
  measured with accuracy.  }. The masses of the planets within our
Solar System are shown with arrows at the top (but are not included in
the data).  The masses of the gas giants fit well with those of
exosolar planets, but the terrestrial planets are all on the low
side. This is most likely due (at least partially) to the difficulty
in finding low-mass planets.  The masses of the exoplanets are
strongly biased towards high--mass and short--period planets. Kepler
has shown us that small planets are very common but the mass
measurement of small mass planets is difficult and thus currently they
appear to be rare.

There are two peaks in the data, the first of which is at a mass
between that of the Earth and that of Uranus, at around $0.01\,M_{\rm
  J}$, where $M_{\rm J}$ is the mass of Jupiter. Planets with a mass
in the range of $1\,M_\oplus$ to $10\,M_\oplus$ (where $M_\oplus$ is
the mass of the Earth) are known as super-Earths
\citep[e.g.][]{Valencia2007}. Our Solar System does not contain any
super-Earths thus in that sense it is somewhat unusual.  We discuss
this further in subsection~\ref{superearth}. The second peak is around
the mass of Jupiter.

We fit the exoplanet mass data with a binormal probability density
function (PDF)
\begin{equation}
P(z)\,dz \propto \frac{1}{\sigma_1} {\rm e}^{\frac{(z-\mu_1)^2}{\sigma_1^2}}+\frac{w_2}{\sigma_2} {\rm e}^{\frac{(z-\mu_2)^2}{\sigma_2^2}},
\label{prob}
\end{equation}
where $z=\log_{10}(m)$.  With a Kolmogorov-Smirnov (KS) test we find
the best fitting parameters to be $\mu_1=-1.80$, $\sigma_1=0.28$,
$w_2=0.69$, $\mu_2=0.12$, $\sigma_2=0.56$ and we show the PDF as the
solid line in Fig.~\ref{mass}.  With this distribution, we find that
Jupiter is very typical at only $-0.26\sigma$ from the higher-mass
peak, while Saturn is at $-1.3\sigma$. The terrestrial planets are
hard to compare because there are so few data points for those small
masses. However, Neptune lies at $1.89\sigma$ and Uranus at
$1.57\sigma$ from the lower-mass peak.

Gas giants are thought to form outside of the snow line radius in the
protoplanetary disk where there is more solid material available to
form massive planets
\citep[e.g.][]{Pollack1996,MartinandLivio2012,MartinandLivio2013snow}.
The surface density of the protoplanetary disk decreases with
increasing distance from the star and the timescale to form a planet
increases with radius. Thus, lower mass gas giants could
preferentially form farther away from the star.  Low mass and large
orbital radius planets are certainly harder to detect than those with
higher mass and lower orbital radius. This can explain the lack of
observed small mass planets, but also perhaps the dip in the observed
distribution. For example, if Uranus and Neptune were around another
star, at their large orbital radii, they would also be difficult to
detect. The only method that could currently detect them is direct
imaging. However, the smallest mass planet that has been found with
this method is Formaulhaut~b that has an approximate mass of $\lesssim
2\,\rm M_{\rm J}$ \citep{Currie2012}.  It therefore remains a
possibility that the double peaked mass distribution is solely the
result of selection effects.

\subsection{Planet Densities}

\begin{figure} 
\begin{centering} 
\includegraphics[width=7.5cm]{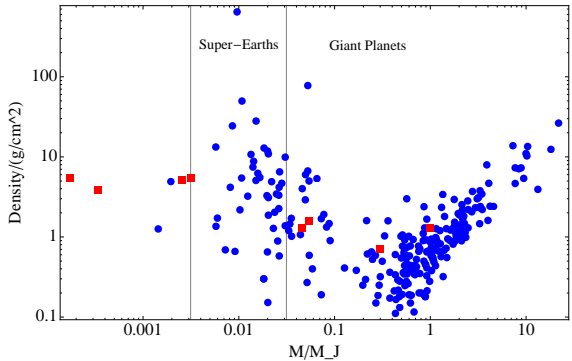} 
\end{centering} 
\caption{Densities of planets as a function of their mass. The blue 
  circles show the exoplanets (total number of 287) and the red 
  squares the planets in our Solar System. The vertical lines shows 
  range of planets that are considered to be super--Earths, the lower 
  limit is the mass of the Earth and the upper limit is the lower 
  limit to the mass of a giant planet at $10\,M_\oplus$. } 
\label{density} 
\end{figure}

Fig.~\ref{density} shows the approximate densities of the observed
planets as a function of their mass. The exoplanets are shown in blue
and the planets in our Solar System in red.  While the lower mass
exoplanets have a large range in their density for a given mass
\citep[see also][]{Wolfgang2012,Marcy2014,Howe2014,Knutson2014}, the
giant planets show a clear correlation of increasing density with
mass. Thus, the super-Earths may have a wide range of compositions
\citep{Valencia2007}. Despite the large spread in the data for the low
mass planets, there appears to be a trend of decreasing density with
increasing mass. This could be attributed qualitatively to the peak in
the theoretical radius against mass of a planet \citep[see
  e.g.][]{zapolsky1969, Seager2007, Fortney2007, Mordasini2012}.

More recently, it has been suggested that the super-Earths (with radii
in the range ${1\,\rm R_\oplus} <R<{4\,\rm R_\oplus}$) show two trends
separated by a critical planet radius. The smallest planets increase
in density with radius while those that are larger decrease suggesting
that the larger planets have a large amount of volatiles on a rocky
core. There is some uncertainty over the value of the critical radius,
as estimates range from about $1.5\,R_\oplus$ to $2\,\rm R_\oplus$
\cite{Petigura2013, Lopez2014, Weiss2014, Marcy2014b}, if it exists at
all \citep{Morton2014}.

The data suggest that the densities of the giant planets within our
Solar System are very typical of those of observed exoplanets.  The
masses of the terrestrial planets in our Solar System are on the edge
of our current sensitivity and thus it is hard to draw any conclusions
about their densities. However, recently, \cite{Dressing2015b} found
that the Earth (and Venus) can be modelled with the same ratio of iron
to magnesium silicate as the low mass exoplanets observed and thus the
Earth may not be special in this respect.

\subsection{Lack of a super-Earth}
\label{superearth}

It is interesting to examine whether our Solar System's lack of a
super-Earth is truly unusual.  There have been several attempts to
calculate an occurrence rate for super-Earths taking into account the
selection biases.  The results for RV observations predict an
occurrence rate in the range $10-20\,\%$ in the period range $P_{\rm
  b}<50\,\rm d$ \citep{Howard2010,Mayor2011}. The transiting planet
observations imply a range of occurrence rates that is at most as high
as $50\,\%$ in the period range $P_{\rm b}<85\,\rm d$
\citep{Fressin2013}. More recently \cite{Burke2015} examined the
Kepler sample for planets with radii in the range
$0.75<R<2.5\,R_\oplus$ with orbital periods in the range 50 to
$300\,\rm d$ and found an occurrence rate of $77\%$. Although these
results include Earth-size planets and some of the periods are longer
than that of Mercury, the occurrence rate increases with short orbital
periods, making the existence of close-in planets more likely. The
high occurrence rate of these types of planets offers perhaps the
strongest argument against the Solar System being very common, but
even that does not necessarily make it extremely rare.  Typically,
systems that have an observed super-Earth, have more than one and this
is theoretically expected if the planets form by mergers of inwardly
migrating cores \citep[e.g.][]{Terquem2007, Cossou2014}.

It is possible that the presence of a super-Earth can affect
terrestrial planet formation. Many of the super-Earths observed are at
small radial locations, where theoretically they could not have formed
\citep[e.g.][]{Raymond2014,Schlichting2014}. \cite{Izidoro2014} found
that if a super-Earth migrates sufficiently slowly through the
habitable zone (defined as the radial range of distances from the star
at which a rocky planet can maintain liquid water on its surface) then
any terrestrial planet that later forms there would be volatile-rich
and not very Earth-like.

In conclusion, the masses and densities of the planets of our
  Solar System appear to be very typical of those of
  exoplanets. However, the lack of a planet with a mass in the range
  $1-10\,M_\oplus$, a super-Earth, makes the Solar System appear
  somewhat special.
 
\section{Discussion} 
\label{discussion}

\begin{figure} 
\begin{centering} 
\includegraphics[width=8.4cm]{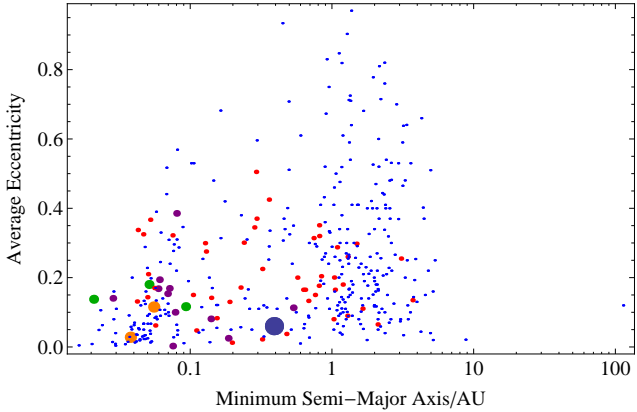} 
\end{centering} 
\caption{The average eccentricity in a planetary system vs the
  smallest planet semi-major axis. The size of the point is
  proportional to the number of planets in the system.  The small blue
  points have 1 planet, the red points have 2 planets, the purple have
  3, the green have 4, the orange have 5 and the Solar System with 8
  is shown in the large blue point. The Solar System has an average
  eccentricity of $0.06$ and Mercury, being the inner most planet, at
  $0.39\,AU$. }
\label{average} 
\end{figure}

Generally, the physical properties of the planets in our Solar System
are quite typical when compared to those of the observed exoplanets,
although the lack of a super--Earth is unusual. The orbital
properties, however, may be somewhat special and perhaps more
conducive to life.  Low eccentricity planets have a more stable
temperature throughout the orbit and therefore may be more likely to
host life \citep[e.g.][]{Williams2002,Gaidos2004}. Furthermore,
planetary systems with a low average eccentricity are more likely to
have long term dynamical stability. For example, the terrestrial
planets in our Solar System are expected to be dynamically stable at
least until the Sun becomes a red giant and engulfs the inner planets
\citep[e.g.][]{Laughlin2009}.

There are a few other factors that could, in principle at least, make
our Solar System special with respect to the emergence of life. First
we can consider the age. The current age of our Sun is about half the
age of the disk of our Galaxy, and also half of the Sun's total
lifetime. Thus, we expect that roughly half of the stars in our
Galaxy's disk are older and half are younger than our Sun. This
implies that the age of our Solar System is definitely not
special. Furthermore, \cite{Behroozi2015} considered the planet
formation history of the Milky way and determined that our Solar
System formed close to the median epoch for giant planet
formation. They also found that about $80\%$ of the currently-existing
Earth-like planets were already formed at the time of the Earth's
formation. We also note that the fact that the Solar System contains a
single star does not make it particularly special, since the binary
fraction in the Kepler sample, for example, is about 50\%
\citep{Horchetal2014}.

The presence of terrestrial planets in the habitable zone around their
host star appears to be quite common. For example, \cite{Dressing2015}
examined the Kepler data for M dwarfs and found that for orbital
periods shorter than $50\,\rm d$, the occurrence rate of Earth-size
planets in the habitable zone is around $18 - 27 \%$. This
conservative estimate could be as high as $\sim50\%$ depending on how
the habitable zone is defined. This is consistent with radial velocity
surveys that find $0.41$ potentially habitable planets per M dwarf
\citep{Bonfils2013}.  Thus, an Earth-size planet in a habitable zone
is not uncommon.

A habitable planet may require a large moon which in turn may require
an asteroid collision \citep{Canup2001}. Thus, systems which contain
an asteroid belt may be more conducive to initiating life. However,
habitability may be sensitive to the size of the asteroid belt
\citep{MartinandLivio2013asteroids}. As we have noted in the
Introduction, asteroid belts could be a common feature of planetary
systems \citep{Chen2014}.

The metallicity of a protoplanetary disk (and hence the host star)
determines the structure of a planetary system that forms
\citep[e.g.][]{Buchhave2014,Wang2015}. The higher the metallicity of a
star, the more giant planets that are observed
\citep[e.g.][]{Fischer2005,Sousa2011, Gonzalez2014,
  Reffert2015}. However, the correlation for lower mass planets is
unclear \citep{Buchhave2012,Mayor2014}. Planets with radii less than
four times that of the Earth are observed around stars with a wide
range of metallicities.  However, the average metallicity of stars
hosting small planets ($R_{\rm p} < 1.7 R_\oplus$) in
\cite{Buchhave2014} is very close to solar. Although such planets can
form at a wide range of metallicities, the fact that the average
metallicity of the small planets is solar may not be a coincidence.
Thus, while the metallicity of our Solar System may not be especially
promotive to the formation of a habitable planet, it's unclear whether
the Solar System is special or not in this respect.

The variability of our Sun has been compared to the activity of stars
in the Kepler sample with conflicting conclusions. \cite{Basri2010} and
\cite{Basri2013} found that the Sun is rather typical with only a
quarter to a third of stars in the Kepler sample being more active than
the Sun. On the other hand, \cite{McQuillan2012} found the Sun to be
relatively quiet with $60\%$ of stars being more active. The
difference in the conclusions stems from choices in defining the
activity level of our Sun and the inclusion of stars with fainter
magnitudes in \cite{McQuillan2012}. However, the studies agree that
the active fraction of stars becomes larger for cooler stars.  M
dwarfs have a fraction of $90\%$ that are more active than the
Sun. Thus, compared to other Sun-like stars, our Sun could be typical,
but compared to cooler stars, our Sun is certainly quiet.

In general, there are  three aspects in which the Solar System
differs most from other observed multi-planet systems. First,
the low mean eccentricity of the planets in the Solar System maybe
somewhat special, although this may be accounted for by selection
effects.  Secondly, there is in the total lack of planets inside
Mercury's orbit. Massive planets migrating through the habitable zone
can change the course of planet formation in that region. Overall,
however, processes that could act to clear the inner part of the Solar
Systems (such as giant planet migration), are believed to be operating
within a non-negligible fraction of the exoplanet systems
\citep[e.g.][]{Batygin2015}.  Third, the lack of super-Earths in
our Solar System is somewhat special and could have allowed the Earth
to become habitable.  A close-in super-Earth could also affect the
dynamical stability of a terrestrial planet in the habitable zone and
this should be investigated in future work.  

We consider the first two of these special parameters in more detail
in Fig.~\ref{average}. For planets with measured eccentricity, we plot
the mean eccentricity of the planetary system and the semi--major axis
of the innermost planet observed within the system. In this parameter
space, the Solar System appears to be somewhat special, but far from
being rare. Although, most of the systems with three or more
planets, do have a planet with an orbital semi-major axis smaller than
that of Mercury, this could be at least partly due to selection
effects. The observed semi--major axis of the innermost planet may be
close to complete but the number of planets in each system is
certainly not. If the innermost planet is very close in, then it is
easier to detect the planets outside of its orbit.  If on the other
hand the innermost planet is farther out (e.g. some of the small blue
and red points in Fig.~6) then additional planets will be difficult to
find (see also discussion in Section~\ref{a}).

There are many factors that may be required in order to form a
habitable planet. When we multiply the probabilities for each
together, we may end up with a small probability for such an
event. However, since we currently do not know which factors are truly
important for life to emerge, such an exercise does not make much
sense. If we consider too many details, clearly the Solar System is
special because all systems are different. However, at the moment we
have not identified any parameter that makes the Solar System so
significantly different that it would make it rare.

\section{Conclusions} 
 
We find that the properties of the planets in our Solar System are not
so significantly special compared to those in exosolar systems to make
the Solar System extremely rare.  The masses and densities are
typical, although the lack of a super--Earth sized planet appears
  to be somewhat unusual.  The orbital locations of our planets seem
to be somewhat special but this is most likely due to selection
effects and the difficulty in finding planets with a small mass or
large orbital period. The mean semi-major axis of observed
  exoplanets is smaller than the distance of Mercury to the Sun.  The
relative depletion in mass of the Solar System's terrestrial region
may be important.  The eccentricities are relatively low compared
  to observed exoplanets, although the observations are biased towards
  finding high eccentricity planets.  The low eccentricity, however,
may be expected for multi-planet systems.  Thus, the two
  characteristics of the Solar System that we find to be most special
  are the lack of super-Earths with orbital periods of days to months
  and the general lack of planets inside of the orbital radius of
  Mercury.

From the perspective of habitability the Solar System does not appear
to be particularly special. If exosolar life happens to be rare it
would probably not be because of simple basic physical parameters, but
because of more subtle processes that are related to the emergence and
evolution of life. Since at the moment we don't know what those might
be, we can allow ourselves to be optimistic about the prospects of
detecting exosolar life. We should make every possible effort to
detect and characterise the atmospheres of a few dozen Earth-size
planets in the habitable zone, in the coming two decades.
 
\section*{Acknowledgements} 
 
We thank the anonymous referee for useful comments. This research has
made use of the Exoplanet Orbit Database and the Exoplanet Data
Explorer at exoplanets.org.

\bibliographystyle{mn2e}
\small  
%\bibliography{../../../../tex/martin}

\label{lastpage} 
\end{document}